\documentstyle[aps,epsfig]{revtex}
\def\be{\begin{equation}}
\def\ee{\end{equation}}
\begin{document}
 \title{Calculation of reduced density matrices from correlation functions}
 \author{
Ingo Peschel\\
{\small Fachbereich Physik, Freie Universit\"at Berlin,} \\
{\small Arnimallee 14, D-14195 Berlin, Germany}}
 \maketitle
 \begin{abstract}
 It is shown that for solvable fermionic and bosonic lattice
 systems, the reduced density matrices can be determined 
 from the properties of the correlation functions. This
 provides the simplest way to these quantities which
 are used in the density-matrix renormalization group method.

\end{abstract}
  
\vspace{2cm}
 
Reduced density matrices for solvable fermionic and bosonic lattice
models have been studied in recent years because such operators 
play a central role in the density-matrix renormalization group (DMRG) method 
\cite{White92,White93,DMRG}. 
In contrast to the quantities
used in other cases, they refer to a subset of $\it sites$, not to
a subset of particles. It has been found that they have exponential
form $exp(-{\cal H})$, where $\cal H$ is again a solvable fermionic or
bosonic operator, confined to the chosen subsystem 
\cite{Peschel/Chung99,Chung/Peschel00,Chung/Peschel01}. 
This was derived by starting from the 
total density matrix (usually for the ground state)
and integrating out the degrees of freedom outside
the subsystem. In the case of fermions, this can be done using Grassmann
variables. The procedure is straightforward, but also somewhat tedious. 
However, it was noted recently that for a hopping
model the final result involves only the one-fermion correlation functions
of the system \cite{Cheong/Henley02}. In the following it is shown that one
can go one step further and 
base the considerations completely on correlation functions.  
The density matrices then follow in a very simple and 
transparent way. \\

Consider first a system of free fermions hopping between lattice sites. The 
corresponding Hamiltonian has the general form
 \be
  \hat{H}=-\sum_{n,m} \hat{t}_{n,m} c_n^{\dagger} c_m 
   \label{eqn:hop}
 \ee
where 
the "hat" denotes quantities of the total system. This Hamiltonian has
Slater determinants as eigenstates. Let  $\mid\Psi>$  be such a state and
 \be
  \hat{C}_{n,m} = < c_n^{\dagger} c_m >
  \label{eqn:cf}
 \ee
the one-particle function in this state. The  $\hat{C}_{n,m}$ form a Hermitian matrix $\hat{C}$.
Because  $\mid\Psi>$ is a determinant, all the higher correlation functions can
be expressed by $\hat{C}$, e. g.
 \be
  <c_n^{\dagger}c_m^{\dagger}c_kc_l>=<c_n^{\dagger}c_l><c_m^{\dagger}c_k>-
   <c_n^{\dagger}c_k><c_m^{\dagger}c_l>
   \label{eqn:fac}
  \ee
Now consider a subsystem of $M$ sites for which the notation $i,j$ will be used. 
By definition, the reduced density matrix $\rho$ reproduces all expectation values 
in the subsystem. Therefore the one-particle function is 
 \be
  C_{i,j}= tr(\rho c_i^{\dagger} c_j)
  \label{eqn:rho1}
 \ee
and the higher functions must factorize as in (\ref{eqn:fac}
).  According to Wick's
theorem, this property holds
if $\rho$ is the exponential of a free-fermion operator \cite{Gaudin60}. Thus one can write
 \be
   \rho= {\cal K} \exp{({-\cal H} )}
   \label{eqn:rho2}
 \ee
where $\cal K$ is a normalization constant and
 \be
  {\cal H}= \sum_{i,j} H_{i,j} c_i^{\dagger} c_j
  \label{eqn:H1}
 \ee

Let $\phi_k(i)$ be the eigenfunctions of $H$ with eigenvalues $\varepsilon_k$.
Then the transformation to new fermion operators $a_k$
 \be
  c_i=\sum_k  \phi_k(i) a_k
  \label{eqn:trans}
 \ee
diagonalizes $\cal H$ and $\rho$ becomes
 \be
   \rho= {\cal K} \exp{(-\sum_{k=1}^{M} \varepsilon_k a_k^{\dagger} a_k)}
   \label{eqn:rho3}
  \end{equation} 
Using this in (\ref{eqn:rho1}) together with $tr(\rho)=1$ gives
\be
 C_{i,j} = \sum_k \phi^*_k(i) \phi_k(j) \frac{1}{e^{\varepsilon_k}+1}
  \label{eqn:C}
 \ee
On the other hand, $H$ has the representation
 \be
  H_{i,j} = \sum_k \phi_k(i) \phi^*_k(j)\; \varepsilon_k
  \label{eqn:H}
 \ee
Therefore the eigenvalues of the two matrices are related by
 \be
  \zeta_k = (e^{\varepsilon_k}+1)^{-1}
  \label{eqn:zeps}
 \ee
 and in matrix form, with the prime denoting the transpose
 \be
    H' = ln{[(1-C)/C]}
  \label{eqn:HC}
 \ee
This is the formula found in \cite{Cheong/Henley02}. Due to its form,
$\rho$ is completely determined by the $M \times M$ matrix $C$. 
One should note
that $\it any$ one-particle correlation function can be expressed in such
a way through a proper free-fermion operator. The only condition is that the
eigenvalues $\zeta_k$ of $C$ lie between 0 and 1 and this is always the case,
since they can be written in the form $<a_k^{\dagger} a_k>$ 
with new fermion operators \cite{Yang62}. However, for
a state which is not a Slater determinant, the free-fermion density matrix
found above would in general give wrong results for other expectation values.\\

These considerations can be extended to systems with pair creation and
annihilation which can be diagonalized by a Bogoliubov transformation.
The eigenstates are then Slater determinants in the new Fermi operators.
In such a state, "anomalous" correlation functions
\be
  \hat{ F}_{n,m} = < c_n^{\dagger} c_m^{\dagger} >
  \label{eqn:cfF}
 \ee
exist which also occur in the factorization equations. Thus (\ref{eqn:fac}) is changed into
 \be
  <c_n^{\dagger}c_m^{\dagger}c_kc_l>=<c_n^{\dagger}c_l><c_m^{\dagger}c_k>-
   <c_n^{\dagger}c_k><c_m^{\dagger}c_l>+<c_n^{\dagger}c_m^{\dagger}><c_kc_l>.
   \label{eqn:fac1}
  \ee
To reproduce this, $\rho$ has to be an exponential with an operator $\cal H$
which also contains pair creation and annihilation processes
 \be
  {\cal H}\;=\; \sum_{i,j} [ c_i^{\dagger} A_{ij} c_j+
         \frac{1}{2} (c_i^{\dagger} B_{ij} c_j^{\dagger} + h.c.) ]
       \label{eqn:H2}
 \ee
Since now two matrices appear in $\cal H$, one needs additional
input, which is provided by the correlation functions $F_{i,j}$.
By following the usual diagonalization procedure for $\cal H$ \cite{Lieb61}
and calculating $C$ and $F$ one can then show that
 \be
  [(C-1/2-F)(C-1/2+F)]_{i,j}= \frac {1}{4} \sum_k \phi_k(i) \phi^*_k(j) th^2({\varepsilon_k}/2)
  \label{eqn:CF}
 \ee
where the $\phi_k(i)$ are the orthonormal eigenfunctions of $(A-B)(A+B)$
and the $\varepsilon_k$ 
are again the single-particle eigenvalues of $\cal H$. Thus one can find the
$\varepsilon_k$ from the eigenvalues of the
matrix on the left hand side of (\ref{eqn:CF}). This matrix can be written as
$KK^{\dagger}/4$ where $K/2=(C-1/2-F)$, since  $F$ is anti-Hermitian.
 For $F=0$, the result is equivalent to (\ref{eqn:C}). If one turns a hopping
 model into one with pair terms via a particle-hole transformation, 
 (\ref{eqn:C}) goes over into (\ref{eqn:CF}). To make
 contact with the treatment in \cite{Chung/Peschel01}, one first
 relates the matrix $\hat{G}$, used there to write an even-parity eigenstate in the form
  \begin{equation} 
   \mid\Psi> \;={\cal C} \exp{\{\frac{1}{2}
               \sum_{n,m} \hat{G}_{nm} c_n^{\dagger} c_m^{\dagger}\}} \mid 0>
, 
   \label{eqn:expform}
  \end{equation}       
 where $\mid 0>$ is the vacuum of the $c_n$, to the quantities appearing here.
 For the ground state, $\hat{G}$ connects the two sets of functions $\hat{\phi_p}$
 and $\hat{\psi_p}$ arising in the diagonalization of the Hamiltonian. The same holds
 for $\hat{K}/2=(\hat{C}-1/2-\hat{F})$ and 
 one finds that $\hat{G}=(\hat{K}-1)/(\hat{K}+1)$. 
 Using this, one can show that the
 matrix in (A9) of \cite{Chung/Peschel01} equals $2(1+KK^{\dagger})/(1-KK^{\dagger})$
 \cite{erratum}. Therefore the eigenvalue equation used in \cite{Chung/Peschel01} to
 determine $2ch\varepsilon_k$ is an alternative version of the relation (\ref{eqn:CF})
 and both approaches are fully consistent.\\

  In a similar way, one can treat systems of coupled harmonic oscillators.
In this case, it is convenient to consider the correlation functions of
positions and momenta
 \be
  \hat{X}_{n,m} = <x_nx_m>,\;\hat{P}_{n,m} = <p_np_m>
  \label{eqn:XP}
 \ee
In the ground state, which is a Gaussian in the coordinates, one then
has factorization formulae like
 \be
  <x_nx_mx_kx_l>=<x_nx_m><x_kx_l> + <x_nx_k><x_mx_l>  +<x_nx_l><x_mx_k>
  \label{eqn:perm}
 \ee
which are non-trivial even if all indices are equal. They hold also if
the expectation values are calculated with an exponential operator
quadratic in the $x$ and $p$. Therefore $\rho$ has the form (\ref{eqn:rho2}) with
\be
  {\cal H}= \frac{1}{2}\sum_{i,j}\; [T_{i,j}p_ip_j +V_{i,j}x_ix_j]
  \label{eqn:H3}
 \ee
The diagonal form is again (\ref{eqn:rho3}) but with bosonic operators and the 
$\varepsilon_k$ follow from the eigenvalues $\nu_k^2$ 
of the matrix $XP$ via
 \be
   cth{(\varepsilon_k/2)}=\nu_k/2
  \label{eqn:bos}
 \ee
If the subsystem is a single
oscillator $i$, there is only one $\nu$ given by $<x_i^2><p_i^2>$ .
For a homogeneous system, this can also be expressed through 
the frequency moments of the normal modes as $<1/\omega><\omega>$. 
The general equations
are again equivalent to those obtained previously in \cite{Chung/Peschel00}.\\

This shows that the way to reduced density matrices associated with eigenstates of
solvable fermionic or bosonic systems can be shortened considerably. 
The results are also valid for systems at
finite temperature as considered in \cite{Cheong/Henley02}. In connection with the DMRG, 
the main aim has been to determine the spectra of the $\rho$ and their general features.
Here the present approach helps if the necessary correlation functions have simple
analytic expressions. This is the case for nearest-neighbour hopping on a chain, or on
a square lattice for half filling. Then only the diagonalization of the matrix
$C$ remains. There are some limitations, because if large $\varepsilon_k$ occur, the 
corresponding eigenvalues of $C$ are exponentially close to 0 or 1.
Also the relation of $\rho$ to the corner transfer matrices
of two-dimensional models \cite{Pescheletal99,Peschel/Chung99} is not visible
unless one determines $\cal H$ in its non-diagonal form. Still, one comes very close
to an analytical solution, and working with the correlations can give
additional insight into the nature of the problem.\\

$\it Note$ $ \it added:$  Formulae similar to those given here can also be found in a recent paper 
on entangled quantum states \cite{Vidal02}.


\begin{thebibliography}{99}
\bibitem{erratum} In Ref. [6], Equ.(15) has to read $\alpha=a^{11}-ca^{22}c'$
and $\alpha/2$ should appear in (16)
\end{thebibliography}
\end{document}